\documentclass[conference]{IEEEtran}

\usepackage{graphicx}
\usepackage{epstopdf} 
\usepackage{algorithm}
\usepackage{algorithmic}
\usepackage{booktabs}
\usepackage{amsmath}
\usepackage[colorlinks,linkcolor=black,anchorcolor=black,citecolor=black,bookmarks=false]{hyperref}
\usepackage{amsmath,amssymb}
\usepackage{svg}
\usepackage{cite}

\ifCLASSINFOpdf
\else
\fi
\begin{document}
%

\title{Sparse Channel Estimation for Pixel Antennas: Addressing the Pilot Rank Deficiency}

\author{Yiting~Chen$^1$, Yumeng Zhang$^2$, and Hongyu~Li$^3$\\
\IEEEauthorblockA{$^1$ School of Information Science and Engineering, Southeast University, Nanjing, China}
\IEEEauthorblockA{$^2$ Dept. of Electronic and Electrical Engineering, Hong Kong University of Science and Technology}
\IEEEauthorblockA{$^3$ Internet of Things Thrust, Hong Kong University of Science and Technology (Guangzhou)\\ Corresponding Author E-mail: \texttt{hongyuli@hkust-gz.edu.cn} }}

\maketitle

\begin{abstract}
Composed of multiple interconnected pixels controlled by on/off RF switches, the pixel antenna can generate reconfigurable radiation patterns that can be further exploited to construct diverse pilot sequences for effective channel estimation. However, such pilot sequences inherently have rank deficiency, making it difficult to effectively and efficiently acquire the full channel state information (CSI) across all available radiation patterns. To tackle this difficulty, we consider a sparse environment with a limited number of propagation paths for a pixel antenna system, where a user equipped with a pixel antenna transmits only a limited number of pilots to recover the CSI under all radiation patterns. The proposed algorithm exploits the limited number of propagation paths that are invariant with the pixel antenna patterns, and then formulates the full channel estimation as a sparse recovery problem in the angular domain solved by Generalized Approximate Message Passing (GAMP). Moreover, to mitigate the rank deficiency of pilot sequences, we additionally incorporate a Multipath Matching Pursuit (MMP) algorithm for robust initialization. The overall proposed scheme, termed MMP-GAMP, achieves higher estimation accuracy than other algorithm baselines, while requiring lower pilot overhead.
\end{abstract}


\IEEEpeerreviewmaketitle

\section{Introduction}
The multiple-input multiple-output (MIMO) technology is a cornerstone in modern and future wireless communications, with research increasingly shifting toward electromagnetic information theory to exploit fundamental limits of the electromagnetic space~\cite{10536068}. One important limitation is that conventional MIMO systems still rely on antennas with fixed configurations, which excludes the antenna hardware itself from system-level optimization. This fixed nature of antennas inherently decouples the antenna response from the dynamic channel states, limiting wireless communication performance.

Pixel antennas have emerged as a revolutionary reconfigurable antenna technology that can break through the isolation between antenna response and dynamic channel states, and introduce additional degrees of freedom for wave manipulation~\cite{6236033,9785489,9743796}. The pixel antenna discretizes a continuous radiation surface into smaller pixel grids interconnected by radio frequency (RF) switches, thus enabling reconfigurable antenna characteristics in radiation patterns, operating frequencies, and polarizations~\cite{9743796}. Recently, pixel antennas have been applied in various wireless communication systems to replace conventional antennas, such as single-antenna and MIMO communications~\cite{shen2026antenna,han2026exploiting}, and multi-user communications~\cite{li2026antenna}. Results in \cite{shen2026antenna} and \cite{li2026antenna} demonstrate that pixel antennas can significantly enhance MIMO channel capacity by up to 4.9 dB and double the multi-user sum rate compared to conventional antenna systems. These results highlight the benefit of pixel antennas in improving wireless communications.

It is worth noting that the aforementioned benefits of pixel antennas are achieved based on perfect channel state information (CSI). It is, however, challenging to acquire accurate channel knowledge for pixel antenna systems due to the following two reasons. First, different from conventional antenna systems where the channel is decoupled from antenna radiation patterns and is fixed within the coherence time, the channel in pixel antenna systems is itself reconfigured by the pixel antenna. In this sense, Least Squares (LS) and Linear Minimum Mean Square Error (LMMSE) estimation require prohibitive overhead to obtain the pattern dependent CSI for all possible radiation patterns of pixel antennas. This raises the first fundamental question: \textit{How to extract the pattern-independent channel information that is sufficient to optimize pixel antenna systems?} Second, the accurate channel estimation typically relies on orthogonal pilot sequences. However, due to the hardware limitation, pixel antenna naturally generates rank deficient pilot sequences. This raises the second fundamental question: \textit{How to accurately estimate channel with rank deficient pilot sequences?}

To answer the above questions, we consider sparse recovery assuming a small number of propagation paths and hence features sparsity in the environmental 2D angle domain (angles of arrival (AoA)-angles of departure (AoD)), where path gains are invariant across pixel antenna radiation patterns, thereby reducing the demand for pilot overhead. We then develop a structured sparse channel estimation framework for a pixel antenna empowered system, consisting of a base station (BS) equipped with multiple conventional antennas and a user equipped with a pixel antenna. It is worth noting that exploiting this joint AoA-AoD sparsity with a single RF chain is uniquely facilitated by the pixel antenna. By dynamically reconfiguring its radiation pattern across successive pilot transmissions, the pixel antenna generates the necessary `virtual' spatial diversity to resolve angular  paths from departure, a capability typically requiring multiple RF chains in conventional systems. The main contributions of this work are summarized as follows.
\addtolength{\topmargin}{0.02 in}
\textit{First,} we carefully select pixel antenna radiation patterns to construct an effective sensing matrix for channel estimation that avoids extremely high spatial correlation. \textit{Second,} we derive an uplink channel model for a pixel antenna empowered system and formulate the AoA-AoD angular domain channel estimation problem. \textit{Third,} we employ the Generalized Approximate Message Passing (GAMP) algorithm for sparse channel recovery in the angular domain. Furthermore, to enhance robustness against rank-deficient sensing matrices, we introduce a multipath matching pursuit (MMP) based initialization to provide a reliable prior for the GAMP algorithm. \textit{Fourth,} we provide simulation results, showing that the proposed method achieves higher channel estimation accuracy and stronger robustness with limited pilot resources.

\textit{Notation:} Throughout this paper, boldface lowercase and uppercase letters denote vectors and matrices, respectively; $(\cdot)^\mathsf{T}$, $(\cdot)^\mathsf{H}$, and $(\cdot)^{-1}$ denote the transpose, Hermitian transpose, and matrix inverse, respectively; $\mathsf{diag}(\cdot)$ denotes a diagonal matrix formed from its argument; $\|\cdot\|_1$, $\|\cdot\|_2$, and  $\|\cdot\|_\mathrm{F}$ denote $\ell$-1, $\ell$-2, and Frobenius norms, respectively. $\mathcal{CN}(\cdot;\mu,\sigma^2)$ denotes the circularly symmetric complex Gaussian distribution.

 

\section{System Model}
In this section, we introduce the pixel antenna model using multiport network theory, the corresponding channel model from beamspace~\cite{shen2026antenna}, and the signal model to facilitate uplink training for channel estimation.

\subsection{Pixel Antenna Model}

\begin{figure}[t]
    \centering    \includegraphics[width=0.95\columnwidth]{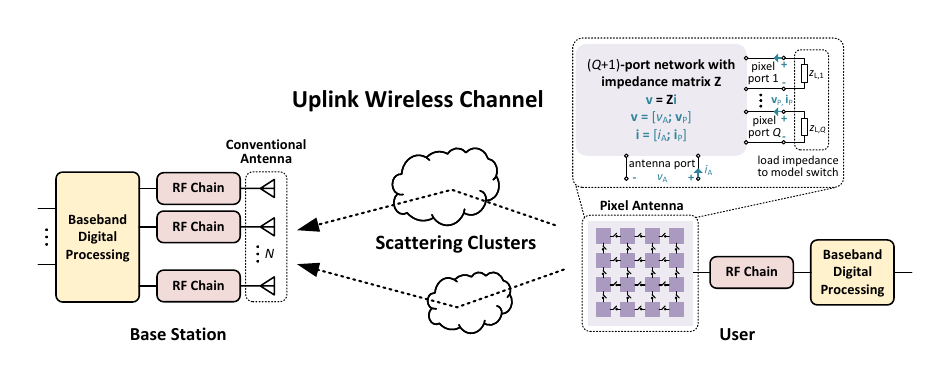}
    \caption{Diagram of a pixel antenna empowered system consisting of a base station with conventional antennas and a user with a pixel antenna.}
    \label{fig:system_model}
\end{figure}

A pixel antenna with $Q$ pixel ports and one antenna port can be modeled as an equivalent $(Q\!+\!1)$-port network, as illustrated in Fig. \ref{fig:system_model}.
Let $\mathbf{Z}\in\mathbb{C}^{(Q+1)\times(Q+1)}$ denote the network impedance matrix partitioned as
\begin{equation}
\mathbf{Z}=\begin{bmatrix}
z_{\mathrm{AA}} & \mathbf{z}_{\mathrm{AP}}\\
\mathbf{z}_{\mathrm{PA}} & \mathbf{Z}_{\mathrm{PP}}
\end{bmatrix},
\end{equation}
where $z_{\mathrm{AA}}\in\mathbb{C}$, $\mathbf{Z}_{\mathrm{PP}}\in\mathbb{C}^{Q\times Q}$, $\mathbf{z}_{\mathrm{AP}}\in\mathbb{C}^{1\times Q}$, and $\mathbf{z}_{\mathrm{PA}} = \
\mathbf{z}_{\mathrm{AP}}^\mathsf{T}$. This impedance matrix links the voltage vector $\mathbf{v} = [v_\mathrm{A},\mathbf{v}_\mathrm{P}^\mathsf{T}]^\mathsf{T}\in\mathbb{C}^{(Q+1)\times 1}$ and the current vector $\mathbf{i} = [i_\mathrm{A},\mathbf{i}_\mathrm{P}^\mathsf{T}]^\mathsf{T}\in\mathbb{C}^{(Q+1)\times 1}$ by $\mathbf{v} = \mathbf{Z}\mathbf{i}$. We further use a binary antenna coder $\mathbf{b} = [b_1,\ldots,b_Q]^\mathsf{T}\in\{0,1\}^{Q\times 1}$ to denote the states of the RF switches, which yields a diagonal load impedance matrix $\mathbf{Z}_\mathrm{L} = \mathsf{diag}(z_{\mathrm{L},1}(b_1),\ldots, z_{\mathrm{L},Q}(b_Q))\in\mathbb{C}^{Q\times Q}$. Specifically, the on state of the switch $q$ is achieved by having $z_{\mathrm{L},q}(b_q) = 0$ with $b_q = 0$ and the off state is achieved by having $z_{\mathrm{L},q}(b_q) = \infty$ with $b_q = 1$. The load impedance of each switch is connected to one pixel port so that we have $\mathbf{v}_{\mathrm{P}}=-\mathbf{Z}_{\mathrm{L}}(\mathbf{b})\mathbf{i}_{\mathrm{P}}$. Accordingly, the currents at the pixel ports can be expressed as
\begin{equation}
\mathbf{i}_{\mathrm{P}}(\mathbf{b})=-\big(\mathbf{Z}_{\mathrm{PP}}+\mathbf{Z}_{\mathrm{L}}(\mathbf{b})\big)^{-1}\mathbf{z}_\mathrm{PA}\, i_\mathrm{A},
\end{equation}
where $i_{\mathrm{A}}\in\mathbb{C}$ is the current at the antenna port.

Let $\mathbf{e}_{\mathrm{A}}=\big[\mathbf{e}_{\mathrm{A},\theta}^\mathsf{T},\mathbf{e}_{\mathrm{A},\phi}^\mathsf{T}\big]^\mathsf{T}\in\mathbb{C}^{2K\times 1}$ denote the open-circuit radiation pattern of the antenna port, where $\mathbf{e}_{\mathrm{A},\theta}\in\mathbb{C}^{K\times 1}$ and $\mathbf{e}_{\mathrm{A},\phi}\in\mathbb{C}^{K\times 1}$ are the $\theta$- and $\phi$-polarized components, respectively. Similarly, let $\mathbf{e}_{\mathrm{P},q}=\big[\mathbf{e}_{\mathrm{P},q,\theta}^{\mathsf{T}},\mathbf{e}_{\mathrm{P},q,\phi}^{\mathsf{T}}\big]^{\mathsf{T}}\in\mathbb{C}^{2K\times 1}$ denote the open-circuit radiation pattern of the $q$-th pixel port ($q=1,\ldots,Q$). Then, the radiation pattern of the pixel antenna coded by the antenna coder $\mathbf{b}$ can be written as
\begin{equation}
\mathbf{e}(\mathbf{b})=\mathbf{E}_{\mathrm{oc}}\mathbf{i}(\mathbf{b}),
\end{equation}
where
$\mathbf{E}_{\mathrm{oc}}=\big[\mathbf{e}_{\mathrm{A}},\mathbf{e}_{\mathrm{P},1},\ldots,\mathbf{e}_{\mathrm{P},Q}\big]\in\mathbb{C}^{2K\times(Q+1)}$
collects the open-circuit radiation patterns of all ports
and $\mathbf{i}(\mathbf{b})=[i_{\mathrm{A}},\,\mathbf{i}_{\mathrm{P}}^\mathsf{T}(\mathbf{b})]^\mathsf{T}\in\mathbb{C}^{(Q+1)\times 1}$ denotes the current vector coded by $\mathbf{b}$. The $\mathbf{e}(\mathbf{b})$ 
is further normalized as $\|\mathbf{e}(\mathbf{b})\|_2 = 1$.

\subsection{Channel Model}
We consider a wireless communication system consisting of a BS having $N$ conventional antennas with fixed configurations and a single-antenna user equipped with a pixel antenna, as illustrated in Fig.~\ref{fig:system_model}. Therefore, the uplink channel coded by the antenna coder can be expressed as~\cite{shen2026antenna}
\begin{equation}
\mathbf{h}(\mathbf{b})=\mathbf{E}_{\mathrm{BS}}^\mathsf{T}\mathbf{H}_{v}\mathbf{e}(\mathbf{b}),
\label{eq:channel_model_compact}
\end{equation}
where $\mathbf{E}_{\mathrm{BS}}=[\mathbf{e}_{\mathrm{BS},1},\ldots,\mathbf{e}_{\mathrm{BS},N}]\in\mathbb{C}^{N\times N}$ collects the radiation patterns of the conventional antennas sampled on $N$ spatial angles, which we assume are unitary for ideally uncorrelated antennas. $\mathbf{H}_{v}\in\mathbb{C}^{N\times 2K}$ denotes the angular-domain virtual channel matrix, and we write
\begin{equation}
\mathbf{H}_{v}=\big[\mathbf{H}_{v,\theta},\mathbf{H}_{v,\phi}\big],
\end{equation}
where $\mathbf{H}_{v,\theta}\in\mathbb{C}^{N\times K}$ and $\mathbf{H}_{v,\phi}\in\mathbb{C}^{N\times K}$ are the virtual channels for the $\theta$- and $\phi$-polarization, respectively.

Note that the matrix $\mathbf{H}_v$ is typically sparse by featuring a limited number of paths. This indicates that estimating only the dominant nonzero entries in $\mathbf{H}_v$ is sufficient for further beamforming design, and that estimating the sparse $\mathbf{H}_v$ is equivalent to obtaining $\mathbf{h}(\mathbf{b})$ for any $\mathbf{b}$.

\subsection{Signal Model}

During the uplink training, the user transmits a sequence of $T$ signals $s_1,\ldots,s_T$ (we assume $s_1 = \ldots = s_T = 1$ without loss of optimality) with power $P$ in $T$ time slots. The BS received signal at time slot $t$ is
\begin{equation}
\mathbf{y}_t=\sqrt{P}\,\mathbf{h}(\mathbf{b}_t)s_t+\mathbf{n}_t
=\sqrt{P}\,\mathbf{E}_\mathrm{BS}^\mathsf{T}\mathbf{H}_{v}\mathbf{e}(\mathbf{b}_t)+\mathbf{n}_t,
\end{equation}
where $\mathbf{e}(\mathbf{b}_t)$ denotes the radiation patten of the pixel antenna at time slot $t$, and $\mathbf{n}_t\sim\mathcal{CN}(\mathbf{0},\sigma^2\mathbf{I})$ denotes the noise.

Stacking $T$ received signals, we have
\begin{align}
\mathbf{Y} \triangleq \big[\mathbf{y}_1,\ldots, \mathbf{y}_T] 
          = \sqrt{P}\,\mathbf{E}_\mathrm{BS}^\mathsf{T}\mathbf{H}_{v}\mathbf{E}(\mathbf{B}) + \mathbf{N},
\label{eq:stacked_received_signal}
\end{align}
where
\begin{equation}
 \mathbf{E}(\mathbf{B})= [\mathbf{e}(\mathbf{b}_{1}),\ldots,\mathbf{e}(\mathbf{b}_{T})],
\end{equation}
with $\mathbf{B} = [\mathbf{b}_{1},\ldots,\mathbf{b}_{T}]$ is a sensing matrix to support the uplink channel estimation and $\mathbf{N} = [\mathbf{n}_1,\ldots,\mathbf{n}_T]$. To facilitate estimation over the sparse $\mathbf{H}_v$, we left-multiply $\mathbf{E}_\mathrm{BS}^*$ to the received signal $\mathbf{Y}$ and obtain 
\begin{equation}
\mathbf{Y}_\mathrm{R}
=\mathbf{E}_\mathrm{BS}^*\mathbf{Y}=\sqrt{P}\,\mathbf{H}_{v}\mathbf{E}(\mathbf{B})+\mathbf{E}_\mathrm{BS}^*\mathbf{N}.
\label{eq:Yr}
\end{equation}

Recall the second fundamental question in Section I that the open-circuit radiation pattern matrix $\mathbf{E}_{\mathrm{oc}}$ embedded in the sensing matrix $\mathbf{E}(\mathbf{B})$ is highly rank deficient, making it difficult to generate sufficient orthogonal radiation patterns (pilots) to construct $\mathbf{E}(\mathbf{B})$. As a solution, we propose an efficient iterative algorithm that handles $\mathbf{H}_v$ estimation with a rank deficient sensing matrix.




\section{A Two-Module Algorithm for Channel Estimation}

\begin{figure*}[t]
    \centering
    \includegraphics[width=0.92\textwidth]{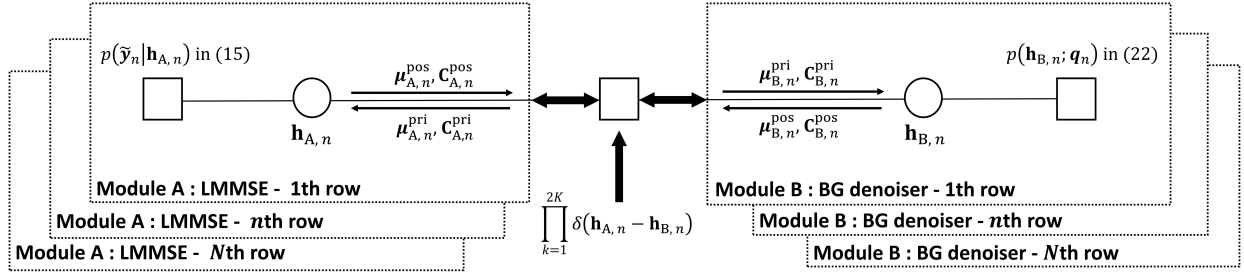}
    \caption{Factor graph of the MMP-GAMP algorithm for channel estimation.}
    \label{fig:algorithm_flow}
\end{figure*}

In this section, we perform channel estimation for $\mathbf{H}_v$ by exploiting its row-wise sparsity. Specifically, we employ the GAMP algorithm as depicted in Fig.~\ref{fig:algorithm_flow}. GAMP iteratively alternates between two modules until convergence: 1) Module A, which processes the measurements $\mathbf{Y}_R$ using LMMSE estimation with prior from Module B, and 2) Module B, which enforces the row-wise sparsity prior via MMSE estimation incorporating prior from Module A. Furthermore, to mitigate the performance degradation caused by highly correlated pilots, we introduce an MMP-based initialization to provide a reliable initialization for the algorithm. The overall algorithm is hence referred to as MMP-GAMP.

To enhance the robustness of GAMP across rank-deficient pilots, we start by constructing the sensing matrix by selecting pilots with the lowest cross-correlation and then applying singular value decomposition (SVD) on the transposed sensing matrix $\mathbf{E}^\mathsf{T}(\mathbf{B})$ as
\begin{equation}
\mathbf{E}^\mathsf{T}(\mathbf{B})=\mathbf{U}\boldsymbol{\Sigma}\mathbf{V}^\mathsf{H},
\end{equation}
where $\mathbf{U}\in\mathbb{C}^{T\times T}$ is a unitary matrix, $\boldsymbol{\Sigma}\in\mathbb{C}^{T\times 2K}$ is a rectangular diagonal matrix composed of singular values, and $\mathbf{V}\in\mathbb{C}^{2K\times 2K}$ is a unitary matrix. Assuming the effective rank $r$ of $\mathbf{E}_{\mathrm{oc}}$ satisfies $r\ll\min\{2K,T\}$, we propose to retain the $r$ dominant singular values of the sensing matrix and obtain
\begin{equation}
\mathbf{E}^\mathsf{T}(\mathbf{B})\approx \mathbf{U}_r\boldsymbol{\Sigma}_r\mathbf{V}_r^\mathsf{H}.
\end{equation}
Substituting this decomposition into \eqref{eq:Yr} and post-multiplying $\mathbf{Y}_\mathrm{R}$ by $\mathbf{U}_r^{*}$ yield the following expression
\begin{equation}
\tilde{\mathbf{Y}}
\triangleq
\mathbf{Y}_\mathrm{R}\mathbf{U}_r^{*}
\approx
\mathbf{H}_v\mathbf{E}^\mathsf{T}+\tilde{\mathbf{N}}_r,
\end{equation}
where $\mathbf{E}\approx\sqrt{P}\,\boldsymbol{\Sigma}_r\mathbf{V}_r^\mathsf{H}\in\mathbb{C}^{r\times 2K}$ is a full-rank sensing matrix for channel estimation, and $\tilde{\mathbf{N}}_r\triangleq\mathbf{E}_\mathrm{BS}^*\mathbf{N}\mathbf{U}_r^{*}$. Since $\mathbf{U}_r$ is semi-unitary, the transformed noise preserves the Gaussian statistics.
We then write the $n$-th row of $\tilde{\mathbf{Y}}$ as \begin{equation}\label{eq:est_y}
\tilde{\mathbf{y}}_{n}
=
\mathbf{E}\mathbf{h}_n
+\tilde{\mathbf{n}}_n,
\end{equation}
where $\mathbf{h}_n = [\mathbf{H}_v^\mathsf{T}]_{:,n}\in\mathbb{C}^{2K\times 1}$ and $\tilde{\mathbf{n}}_n = [\tilde{\mathbf{N}}_r^\mathsf{T}]_{:,n}\in\mathbb{C}^{2K\times 1}$. 

The expression \eqref{eq:est_y} is a standard form that enables channel estimation, which, however, is computationally intractable to simultaneously incorporate the measurements and sparsity prior together. As a solution, we adopt GAMP, whose factor graph is depicted in Fig. \ref{fig:algorithm_flow} and explicitly characterizes the statistical dependencies between the channel variables (the circles) and the underlying probabilistic models (the squares). 

Specifically, in GAMP, we conceptually split $\mathbf{h}_n$ into two identical copies,  denoted by $\mathbf{h}_{\mathrm{A},n}$ and $\mathbf{h}_{\mathrm{B},n}$, corresponding to Modules A and B respectively. Module A incorporates the measurement model and performs LMMSE-based posterior inference, while Module B incorporates the Bernoulli--Gaussian (BG) prior to exploit the row-wise sparsity. To enhance the consistency between two modules, the equality function $\delta(\mathbf{h}_{\mathrm{A},n}-\mathbf{h}_{\mathrm{B},n})$ is introduced in the factor graph, i.e., each module computes its posterior distribution and passes the resulting extrinsic message to the other module as prior information until convergence \cite{7869633,11373228}. 

Below, we will describe the posterior inference in Modules A and B for GAMP in detail.

\subsection{Module A: LMMSE Estimation}
Module A treats the output messages from Module B as a Gaussian prior and refines it using the observed measurements via an LMMSE update. 

Firstly,  we model the prior messages passed from Module B as
\begin{equation}\label{eq:modelA_prior}
{\mathbf{h}}_{\mathrm{A},n}^{\mathrm{pri}} \sim \mathcal{CN}\!\left(\boldsymbol{\mu}_{\mathrm{A},n}^{\mathrm{pri}},\,\mathbf{C}_{\mathrm{A},n}^{\mathrm{pri}}\right),
\end{equation}
where $\boldsymbol{\mu}_{\mathrm{A},n}^{\mathrm{pri}}\in\mathbb{C}^{2K\times 1}$ is the prior mean of $\mathbf{h}_n$ passed from Module~B , and $\mathbf{C}_{\mathrm{A},n}^{\mathrm{pri}}$ is the associated diagonal prior covariance.

Furthermore, given \eqref{eq:est_y}, the likelihood function for $\tilde{\mathbf{y}}_n$ conditioned on $\mathbf{h}_{\mathrm{A},n}$ is
\begin{equation}\label{eq:likelihod}
p(\tilde{\mathbf{y}}_n|\mathbf{h}_{\mathrm{A},n})=\frac{1}{(\pi\sigma^2)^{2K}}\exp\left\{{-\frac{1}{\sigma^2}}\|\tilde{\mathbf{y}}_n-\mathbf{E}\mathbf{h}_{\mathrm{A},n}\|_2^2\right\}.
\end{equation}

After incorporating \eqref{eq:modelA_prior} and \eqref{eq:likelihod} via LMMSE, the resultant posterior inference is the following Gaussian distribution \cite{11373228}
\begin{equation}
{\mathbf{h}}_{\mathrm{A},n}^{\mathrm{pos}}\sim\mathcal{CN}\!\left(\boldsymbol{\mu}_{\mathrm{A},n}^{\mathrm{pos}},\,\mathbf{C}_{\mathrm{A},n}^{\mathrm{pos}}\right),
\end{equation}
with mean and diagonal covariance being
\begin{align}
\boldsymbol{\mu}_{\mathrm{A},n}^{\mathrm{pos}}
&=\mathbf{C}_{\mathrm{A},n}^{\mathrm{pos}}\left((\mathbf{C}_{\mathrm{A},n}^{\mathrm{pri}})^{-1}\boldsymbol{\mu}_{\mathrm{A},n}^{\mathrm{pri}}+
\sigma^{-2}\mathbf{E}^\mathsf{H}\tilde{\mathbf{y}}_n\right),\label{eq:moduleA_mean}\\
\mathbf{C}_{\mathrm{A},n}^{\mathrm{pos}}
&=\left((\mathbf{C}_{\mathrm{A},n}^{\mathrm{pri}}+\sigma^{-2}\mathbf{E}^\mathsf{H}\mathbf{E}\right)^{-1}.\label{eq:moduleA_cov}
\end{align}

Then, GAMP passes the posterior message from Module A to Module B by removing the contribution of the prior from Module B. The resulting extrinsic message from Module A to Module B is given by~\cite{7869633}
\begin{align}
\boldsymbol{\mu}_{\mathrm{B},n}^{\mathrm{pri}}
&=
\mathbf{C}_{\mathrm{B},n}^{\mathrm{pri}}
\left(
(\mathbf{C}_{\mathrm{A},n}^{\mathrm{pos}})^{-1}\boldsymbol{\mu}_{\mathrm{A},n}^{\mathrm{pos}}
-
(\mathbf{C}_{\mathrm{A},n}^{\mathrm{pri}})^{-1}\boldsymbol{\mu}_{\mathrm{A},n}^{\mathrm{pri}}
\right)\label{eq:ModuleA_extrinsic_mean},\\
\mathbf{C}_{\mathrm{B},n}^{\mathrm{pri}}
&=\left((\mathbf{C}_{\mathrm{A},n}^{\mathrm{pos}})^{-1}-(\mathbf{C}_{\mathrm{A},n}^{\mathrm{pri}})^{-1}\right)^{-1}.\label{eq:ModuleA_extrinsic_cov}
\end{align}

\begin{algorithm}[t]
	\caption{MMP-GAMP Algorithm}
	\label{alg:row_wise_turbo_em_bg}
	\begin{algorithmic}[1]
		\REQUIRE Received signal $\tilde{\mathbf{Y}}=[{\tilde{\mathbf{y}}_1},\dots,{\tilde{\mathbf{y}}_N}]^\mathsf{T}$, sensing matrix $\mathbf{E}$, noise variance $\sigma^2$. 
		\ENSURE Row-wise estimate $\hat{\mathbf{H}}_v=[\hat{\mathbf{h}}_1,\dots,\hat{\mathbf{h}}_N]^\mathsf{T}$.
        \STATE \textbf{0) Initialization:} Set $\{\boldsymbol{\mu}_{\mathrm{A},n}^{\mathrm{pri}},\mathbf{C}_{\mathrm{A},n}^{\mathrm{pri}}\}_{n=1}^{N}$ by \eqref{eq:MMP_mean} and \eqref{eq:MMP_cov} and $\{\lambda_n,\theta_n,\phi_n\}_{n=1}^N$ by \eqref{eq:MMP_lamda}--\eqref{eq:MMP_phi} in Section III-C.
		\WHILE{no convergence}
			\FOR{$n=1$ to $N$}
				\STATE \textbf{1) Module A (LMMSE):} 
                \STATE Compute $\boldsymbol{\mu}_{\mathrm{A},n}^{\mathrm{pos}}$ by \eqref{eq:moduleA_mean} and $\mathbf{C}_{\mathrm{A},n}^{\mathrm{pos}}$ by \eqref{eq:moduleA_cov}.
				\STATE Form extrinsic messages to set Module B BG prior $\big(\boldsymbol{\mu}_{\mathrm{B},n}^{\mathrm{pri}},\mathbf{C}_{\mathrm{B},n}^{\mathrm{pri}}\big)\leftarrow\big(\boldsymbol{\mu}_{\mathrm{A},n}^{\mathrm{pos}},\mathbf{C}_{\mathrm{A},n}^{\mathrm{pos}}\big)$ with \eqref{eq:ModuleA_extrinsic_mean} and \eqref{eq:ModuleA_extrinsic_cov}.
				\STATE \textbf{2) Module B (GAMP):} 
                \STATE Compute $\boldsymbol{\mu}_{\mathrm{B},n}^{\mathrm{pos}}$ by
                \eqref{eq:ModuleB_mean} and $\mathbf{C}_{\mathrm{B},n}^{\mathrm{pos}}$ by \eqref{eq:ModuleB_cov}.  
				\STATE Form extrinsic messages to set Module A BG prior $\big(\boldsymbol{\mu}_{\mathrm{A},n}^{\mathrm{pri}},\mathbf{C}_{\mathrm{A},n}^{\mathrm{pri}}\big)\leftarrow\big(\boldsymbol{\mu}_{\mathrm{B},n}^{\mathrm{pos}},\mathbf{C}_{\mathrm{B},n}^{\mathrm{pos}}\big)$ with \eqref{eq:ModuleB_extrinsic_mean} and \eqref{eq:ModuleB_extrinsic_cov}.
                \STATE \textbf{3) EM:} Update hyperparameters $\mathbf{q}_n^\star=(\lambda_n^\star,\theta_n^\star,\phi_n^\star)$ 
                by \eqref{eq:1}--\eqref{eq:em_update} in Section III-D. 
            \ENDFOR
		\ENDWHILE
		\RETURN $\hat{\mathbf{H}}_v\leftarrow[\boldsymbol{\mu}_{\mathrm{B},1}^{\mathrm{pos}},\dots,\boldsymbol{\mu}_{\mathrm{B},N}^{\mathrm{pos}}]^\mathsf{T}$.
	\end{algorithmic}
\end{algorithm}

\subsection{Module B: Sparse Denoising via GAMP}
In contrast, Module B, while incorporating prior messages passed from Module A,  exploits the sparsity of the angular-domain channel for its posterior inference. 

Firstly,  for each row index $n$, the message passing from Module A formulates a prior distribution at Module B as follows
\begin{equation}
{\mathbf{h}}^{\mathrm{pri}}_{\mathrm{B},n}
\sim
\mathcal{CN}\!\left(
\boldsymbol{\mu}_{\mathrm{B},n}^{\mathrm{pri}},
\mathbf{C}_{\mathrm{B},n}^{\mathrm{pri}}
\right).
\label{eq:module_b_prior}
\end{equation}

In addition, Module B exploits the sparsity of the angular-domain channel by assuming the following BG prior
\begin{equation}
\begin{aligned}
 p(\mathbf{h}_{\mathrm{B},n};\mathbf{q}_n)
 &= (1-\lambda_n)\,\delta(\mathbf{h}_{\mathrm{B},n})\\
 &\quad +\lambda_n\,\mathcal{CN}\big(\mathbf{h}_{\mathrm{B},n};\theta_n,\phi_n\big),
\end{aligned}
\label{eq:bg_prior}
\end{equation}
where $\mathbf{q}_n \triangleq (\lambda_n,\theta_n,\phi_n)$ denotes the BG hyperparameters with $\lambda_n$ being the sparsity rate, $\theta_n$ and $\phi_n$ being the Gaussian mean and variance of the non-zero component of $n$-th row. This prior means that each coefficient is zero with probability $1-\lambda_n$ or is drawn from a complex Gaussian distribution with probability $\lambda_n$, thereby capturing the row-wise sparsity of $\mathbf{H}_v$.

Incorporating the prior distribution in \eqref{eq:module_b_prior} with the sparsity-captured BG distribution in \eqref{eq:bg_prior}, we approximate the posterior distribution as the following Gaussian distribution \cite{11373228}
\begin{equation}
{\mathbf{h}}^{\mathrm{pos}}_{\mathrm{B},n}
\sim
\mathcal{CN}\!\left(
\boldsymbol{\mu}_{\mathrm{B},n}^{\mathrm{pos}},
\mathbf{C}_{\mathrm{B},n}^{\mathrm{pos}}
\right),
\label{eq:module_b_posterior}
\end{equation}
with the mean and covariance diagonal elements given by
\begin{align}
[\boldsymbol{\mu}_{\mathrm{B},n}^\mathrm{pos}]_k
&= \pi_{n,k}\tau_{n,k}\label{eq:ModuleB_mean},\\
[\mathbf{C}_{\mathrm{B},n}^{\mathrm{pos}}]_{k,k}
&= \pi_{n,k}\big(|\tau_{n,k}|^2+\nu_{n,k}\big)-|\pi_{n,k}\tau_{n,k}|^2, \forall k\label{eq:ModuleB_cov},
\end{align}
where 
\begin{align}
\pi_{n,k}
&\triangleq \frac{1}{1+\left(\frac{\lambda_n}{1-\lambda_n}
\frac{\mathcal{CN}\!\big([\boldsymbol{\mu}_{\mathrm{B},n}^{\mathrm{pri}}]_k;\theta_n,\phi_n+[\mathbf{C}_{\mathrm{B},n}^{\mathrm{pri}}]_{k,k}\big)}
{\mathcal{CN}\!\big([\boldsymbol{\mu}_{\mathrm{B},n}^{\mathrm{pri}}]_k;0,[\mathbf{C}_{\mathrm{B},n}^{\mathrm{pri}}]_{k,k}\big)}\right)^{-1}},\\
\tau_{n,k}
&\triangleq
\frac{\phi_n[\boldsymbol{\mu}_{\mathrm{B},n}^{\mathrm{pri}}]_k+[\mathbf{C}_{\mathrm{B},n}^{\mathrm{pri}}]_{k,k}\theta_n}
{\phi_n+[\mathbf{C}_{\mathrm{B},n}^{\mathrm{pri}}]_{k,k}},\\
\nu_{n,k}
&\triangleq
\frac{[\mathbf{C}_{\mathrm{B},n}^{\mathrm{pri}}]_{k,k}\phi_n}
{[\mathbf{C}_{\mathrm{B},n}^{\mathrm{pri}}]_{k,k}+\phi_n}.
\label{eq:bg_parameters}
\end{align}

Then, GAMP passes the posterior message from Module B to Module A by removing the contribution of the prior from Module A. The resulting extrinsic message from Module B to Module A is given by \cite{7869633}
\begin{align}
\boldsymbol{\mu}_{\mathrm{A},n}^{\mathrm{pri}}
&=
\mathbf{C}_{\mathrm{A},n}^{\mathrm{pri}}
\left(
(\mathbf{C}_{\mathrm{B},n}^{\mathrm{pos}})^{-1}\boldsymbol{\mu}_{\mathrm{B},n}^{\mathrm{pos}}
-
(\mathbf{C}_{\mathrm{B},n}^{\mathrm{pri}})^{-1}\boldsymbol{\mu}_{\mathrm{B},n}^{\mathrm{pri}}
\right),\label{eq:ModuleB_extrinsic_mean}\\
\mathbf{C}_{\mathrm{A},n}^{\mathrm{pri}}
&=
\left((\mathbf{C}_{\mathrm{B},n}^{\mathrm{pos}})^{-1}
-
(\mathbf{C}_{\mathrm{B},n}^{\mathrm{pri}})^{-1}\right)^{-1}.
\label{eq:ModuleB_extrinsic_cov}
\end{align}

\subsection{{MMP-based Prior Initialization}}

We use MMP to initialize both the LMMSE module and the BG denoiser~\cite{6762942}.
For the $n$-th row, MMP generates a series of support sets
$\mathcal{S}_{n,1},\mathcal{S}_{n,2},\ldots,\mathcal{S}_{n,S_n}$,
where each $\mathcal{S}_{n,s}\subseteq\{1,\ldots,2K\}$ is an active support storing indices corresponding to possible nonzero channel entries. 
For robustness, the candidate pool can be augmented with simple correlation-based or orthogonal matching pursuit (OMP)-based seed supports.
Then, for each candidate support $\mathcal{S}_{n,s}$, the score is calculated by
\begin{equation}
    J(\mathcal{S}_{n,s})
    =
    \frac{\|\tilde{\mathbf{y}}_n-\mathbf{E}_{\mathcal{S}_{n,s}}\hat{\mathbf{h}}_{\mathcal{S}_{n,s}}\|_2^2}
         {\|\tilde{\mathbf{y}}_n\|_2^2}
    + \beta_n |\mathcal{S}_{n,s}|,
\end{equation}
where $\mathbf{E}_{\mathcal{S}_{n,s}}$ is obtained by forcing columns of $\mathbf{E}$ with indices in $\mathcal{S}_{n,s}$ to zeros, $\beta_n$ denotes the row-wise support penalty, and the estimate $\hat{\mathbf{h}}_{\mathcal{S}_{n,s}}$ is obtained by a row-wise LMMSE rule
\begin{equation}
    \hat{\mathbf{h}}_{\mathcal{S}_{n,s}}
    =
    v_n^{\mathrm{init}} \mathbf{E}_{\mathcal{S}_{n,s}}^\mathsf{H}
    \left(
    v_n^{\mathrm{init}} \mathbf{E}_{\mathcal{S}_{n,s}}\mathbf{E}_{\mathcal{S}_{n,s}}^\mathsf{H} + \sigma_n^2 \mathbf{I}
    \right)^{-1}
    \tilde{\mathbf{y}}_n,
\end{equation}
where $v_n^{\mathrm{init}}$ is the row-dependent prior variance used during the MMP initialization stage. 
The best support is selected as
\begin{equation}
    \mathcal{S}_n^\star = \arg\min_{\mathcal{S}_{n,s}, \forall s} J(\mathcal{S}_{n,s}),\label{eq:best_support}
\end{equation}
and the corresponding estimate is denoted as $\hat{\mathbf{x}}_{\mathcal{S}_n^\star}$.

\subsubsection{Initialization of the LMMSE prior}

The initial mean and variance for the first LMMSE iteration are set as
\begin{align}
    \boldsymbol{\mu}^\mathrm{pri}_{\mathrm{A},n} &= \hat{\mathbf{h}}_{\mathcal{S}_n^\star}\label{eq:MMP_mean},\\
    \mathbf{C}_{\mathrm{A},n}^\mathrm{pri} &= \frac{1}{T}\|\tilde{\mathbf{y}}_n-\mathbf{E}_{\mathcal{S}_n^\star}\hat{\mathbf{h}}_{\mathcal{S}_n^\star}\|_2^2\mathbf{I}\label{eq:MMP_cov}.
\end{align}

\subsubsection{Initialization of the BG hyperparameters}

The BG hyperparameters are initialized in a row-wise manner from the accepted MMP output of each row. First, the sparsity rate of the $n$-th row is initialized as
\begin{equation}
    \lambda_n
    = \frac{\|\mathbf{E}_{\mathcal{S}_n^\star}\hat{\mathbf{h}}_{\mathcal{S}_n^\star}\|_2^2}{\|\tilde{\mathbf{y}}_n\|_2^2},\label{eq:MMP_lamda}
\end{equation}
which measures the relative energy explained by the MMP reconstruction. 
The corresponding row-wise mean and variance of the active Gaussian component are initialized as
\begin{equation}
    \theta_n = \frac{1}{|\mathcal{S}_n^\star|}\|\hat{\mathbf{h}}_{\mathcal{S}_n^\star}\|_1,\label{eq:MMP_theta}
\end{equation}
\begin{equation}
    \phi_n
    = \frac{1}{|\mathcal{S}_n^\star|}\|\hat{\mathbf{h}}_{\mathcal{S}_n^\star} - \theta_n\mathbf{1}\|_2^2.\label{eq:MMP_phi}
\end{equation}

\subsection{Expectation-Maximization (EM) Update}

To further enhance the robustness of the algorithm and prevent hyper-parameter mismatches that could hinder convergence, this section introduces an EM rule to automatically update the hyper-parameters of the proposed MMP-GAMP algorithm. Specifically, following the framework in~\cite{vila2011expectation}, the hyperparameters are updated as
\begin{align}
\lambda_n^\star
&=
\frac{1}{2K}\sum_{k=1}^{2K}\pi_{n,k},\label{eq:1}\\
\theta_n^\star
&=
\frac{\sum_{k=1}^{2K}\pi_{n,k}\tau_{n,k}}
{\sum_{k=1}^{2K}\pi_{n,k}},\label{eq:2}\\
\phi_n^\star
&=
\frac{\sum_{k=1}^{2K}\pi_{n,k}
\left(
|\tau_{n,k}-\theta_n^\star|^2+\nu_{n,k}
\right)}
{\sum_{k=1}^{2K}\pi_{n,k}}.
\label{eq:em_update}
\end{align}

\section{Performance Evaluation}






\begin{figure}[t]
    \centering

    \includegraphics[width=0.9\columnwidth]{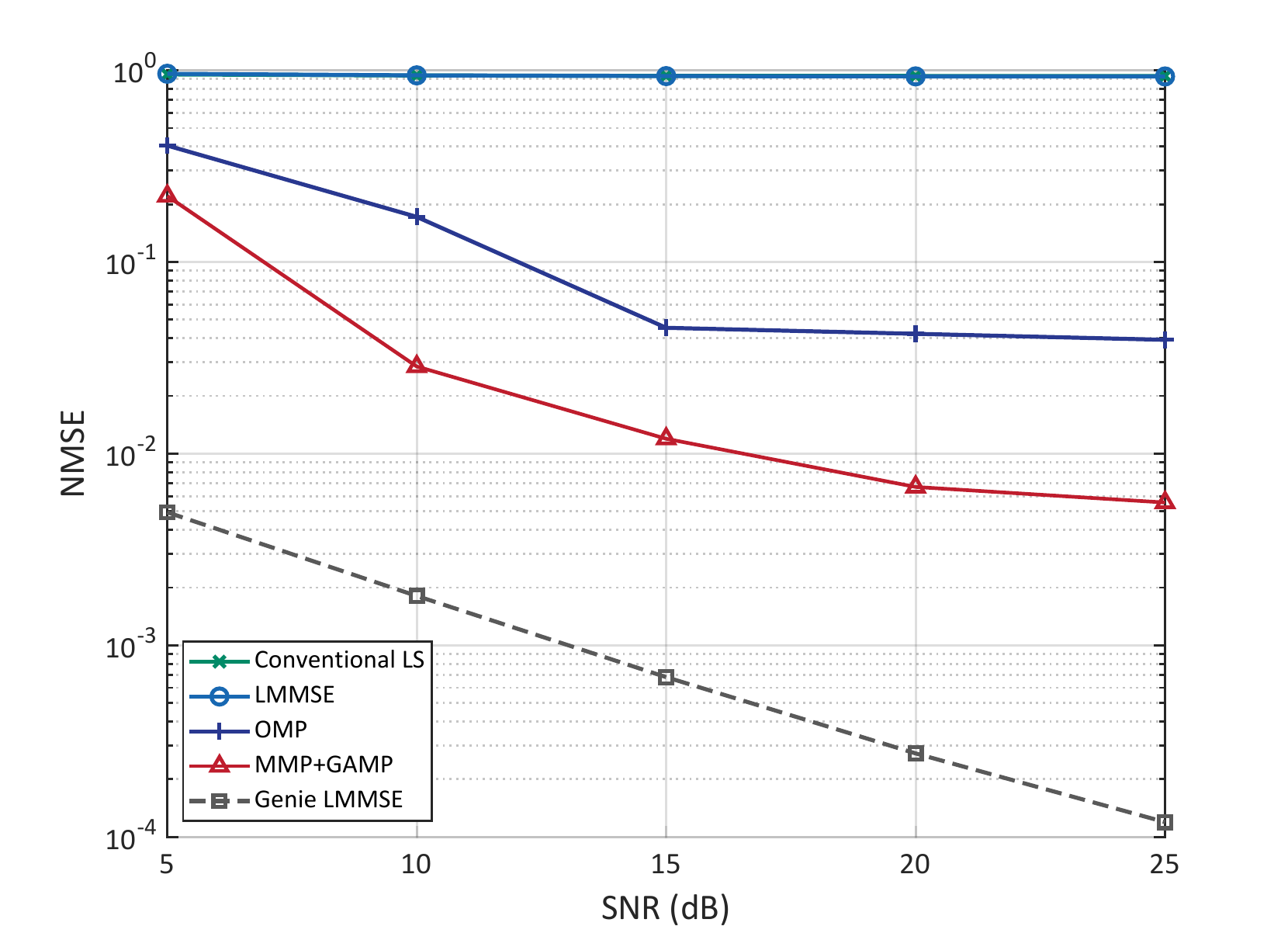}

    \vspace{0.3em}
    {\small (a) $T=10$, $r=8$}

    \vspace{0.8em}

    \includegraphics[width=0.9\columnwidth]{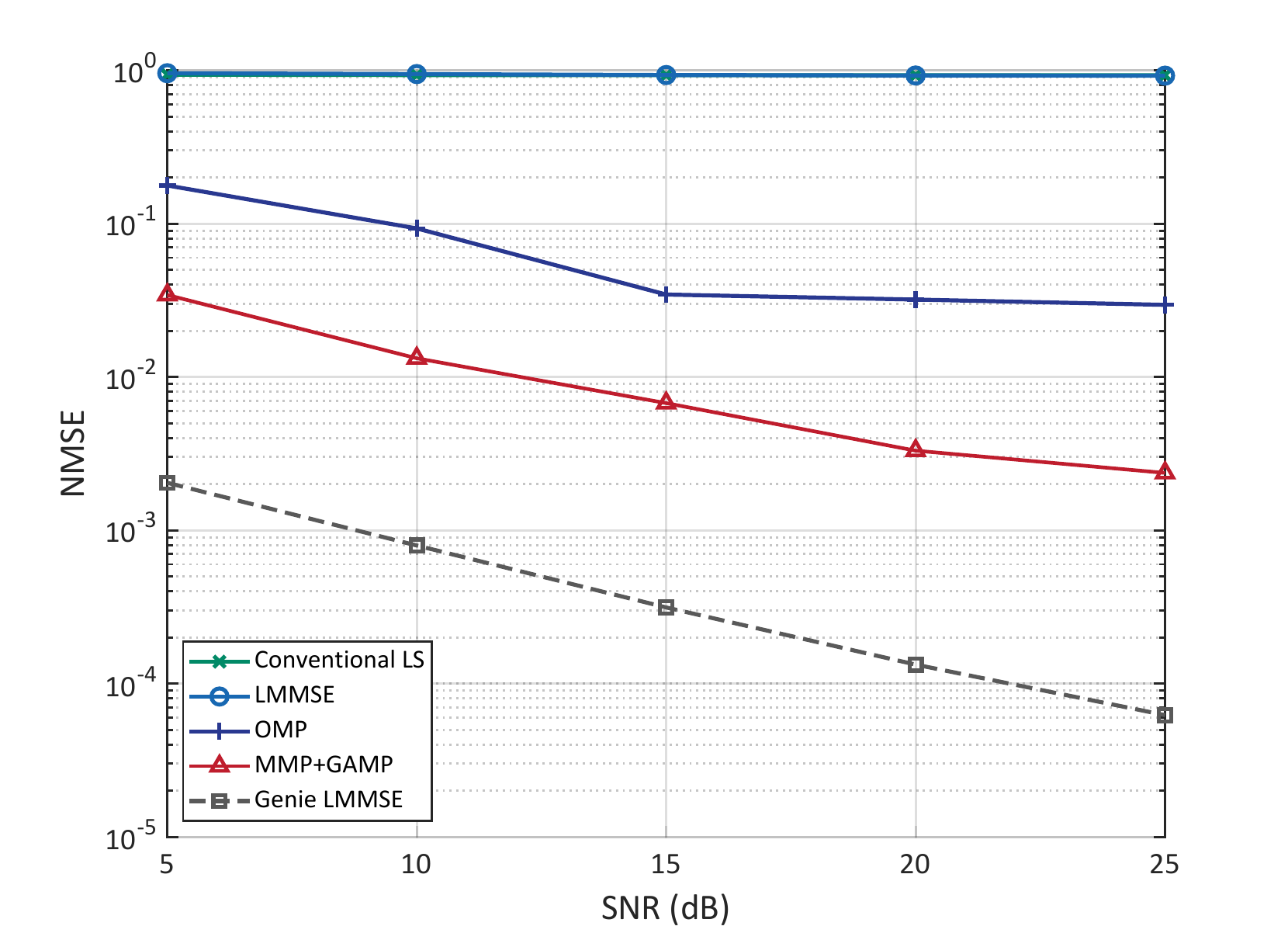}

    \vspace{0.3em}
    {\small (b) $T=30$, $r=9$}

    \caption{NMSE performance of uplink channel estimation using the pixel antenna as a function of signal-to-noise ratio (SNR).}
    \label{fig:umi_los}
\end{figure}

In this section, we evaluate the performance of the proposed MMP-GAMP algorithm in a QuaDRiGa-based UMi line of sight (LoS) scenario~\cite{6758357}. The BS employs $N=16$ conventional antennas, and the user is equipped with a pixel antenna having $Q=39$ pixel ports. The sensing matrix is constructed by selecting pilots with the lowest cross-correlation, which is achieved through random generation of switch states followed by a greedy search. The carrier frequency is $2.4$ GHz. The angular dictionary size is $K=72$. The impedance matrix $\mathbf{Z}$ and the open-circuit radiation pattern matrix $\mathbf{E}_\mathrm{oc}$ are simulated by the Computer Simulation Technology (CST) studio suite. The BS is located at $(0,0,10)$ m and the user is located at $(18,0,1.5)$ m. The spatial-domain channel is projected onto a DFT dictionary, where the two transmit polarizations are stacked over the same AoD grid. To create a sparse propagation environment, we set the Ricean factor to $24$ dB, the mean cross-polarization ratio to $0$ dB, and reduce the delay and angular spreads. The estimation accuracy is measured by the normalized mean square error (NMSE), i.e.,
\begin{equation}
    \mathrm{NMSE}=\frac{\|\mathbf{H}_v-\hat{\mathbf{H}}_v\|_\mathrm{F}^2}{\|\mathbf{H}_v\|_\mathrm{F}^2}.
\end{equation}

To evaluate the effectiveness of the proposed algorithm, we include the following baselines
\begin{itemize}
    \item \textbf{LS}: A fundamental baseline estimator serves as a reference for the rank deficient setting.
    \item \textbf{LMMSE}: A statistical linear baseline without sparsity prior. Specifically, it follows \eqref{eq:moduleA_mean} and \eqref{eq:moduleA_cov} by setting $\boldsymbol{\mu}_{\mathrm{A},n}^{\mathrm{pri}}=\mathbf{0}$ and $\mathbf{C}_{\mathrm{A},n}^{\mathrm{pri}}=0.008\mathbf{I}$.
    \item \textbf{OMP}: A classical greedy compressed-sensing algorithm that exploits the sparsity for channel recovery.
    \item \textbf{Genie LMMSE}: A baseline that achieves the best performance but practically unachievable with oracle support information. Specifically, it follows \eqref{eq:moduleA_mean} and \eqref{eq:moduleA_cov} by setting $\boldsymbol{\mu}_{\mathrm{A},n}^{\mathrm{pri}}=\mathbf{0}$ and using the true support together with the exact per-coefficient prior variance, i.e., $\mathbf{C}_{\mathrm{A},n}^{\mathrm{pri}}=\mathsf{diag}(|\mathbf{h}_{n,\mathcal{S}_{n}^\mathrm{true}}|^2)$ over the true oracle support $\mathcal{S}_{n}^\mathrm{true}$.
\end{itemize}

\begin{figure}[t]
    \centering
    \includegraphics[width=0.98\columnwidth]{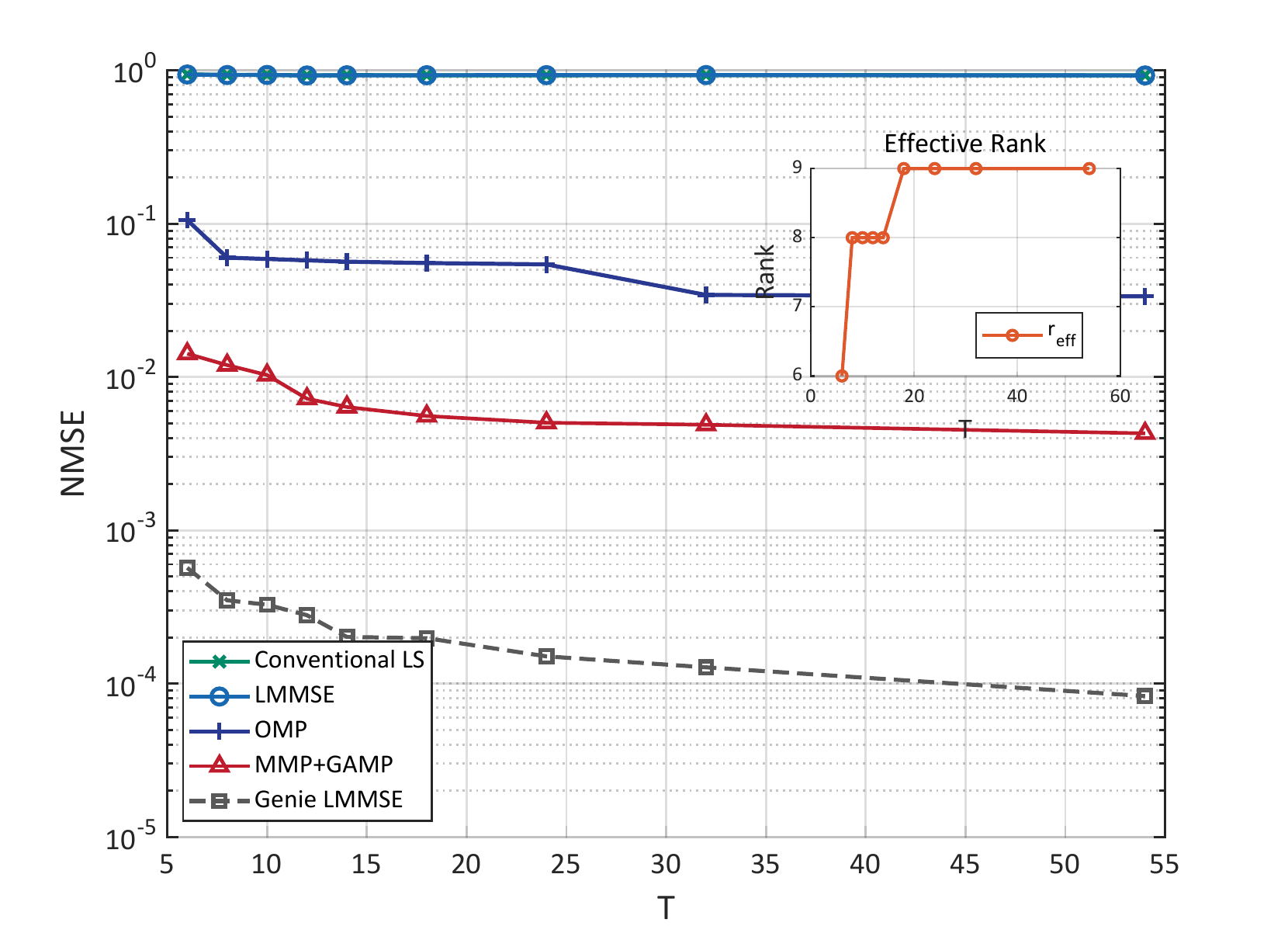}
    \caption{NMSE performance of uplink sparse channel estimation using pixel antenna versus pilot overhead $T$ (SNR $=$ 20 dB). }
    \label{fig:nmse_vs_t}
\end{figure}

Fig.~\ref{fig:umi_los} illustrates the NMSE versus SNR for different channel estimation algorithms under low or high pilot overheads $T$. The LS and LMMSE baselines fail to estimate the virtual channel since they do not exploit the channel sparsity and rely heavily on orthogonal pilot sequences, whereas OMP and MMP-GAMP achieve noticeable improvement by exploiting channel sparsity. Under low pilot overhead, the proposed MMP-GAMP method further reduces the NMSE by around $3$--$5$ dB compared to OMP with relatively high SNR. Under high pilot overhead, benefiting from the improved condition of sensing matrix, MMP-GAMP achieves lower NMSE, e.g., below $10^{-2}$ at SNR $=$ 15 dB. Besides, the Genie-LMMSE baseline remains the best as expected, and the gap between Genie-LMMSE and MMP-GAMP reflects the performance loss of practical support learning without oracle prior knowledge.

Fig.~\ref{fig:nmse_vs_t} shows the NMSE versus pilot overhead $T$ with fixed SNR$=$20 dB. The LS and LMMSE baselines cannot recover the virtual channel regardless of the pilot overhead due to the rank deficiency of the sensing matrix. OMP performs better by exploiting sparsity, but in the low-$T$ regime it cannot reliably identify the channel support and thus requires a larger $T$ to show clear improvement. In contrast, the proposed MMP-GAMP scheme achieves better NMSE performance over the whole range of $T$, yielding about $10$--$14$ dB gain over OMP and more than $20$ dB gain over LS and LMMSE with a relatively large $T$.

\section{Conclusion}
This paper studied sparse uplink channel estimation using pixel antenna with rank deficient pilot sequences and limited pilot overhead. We formulated full CSI recovery as a structured sparse estimation problem in the 2D joint AoA-AoD domain, developed an SVD-aided MMP-GAMP estimator with robust initialization and iterative Bayesian refinement, and demonstrated in a QuaDRiGa-based UMi LoS scenario that the proposed scheme consistently outperforms LS, LMMSE, and OMP under both low and moderate pilot overheads.

\section*{Acknowledgment}

The authors would like to acknowledge Dr. Shanpu Shen with the University of Macau for his professional guidance on modeling the channel empowered by the pixel antenna.







\bibliographystyle{IEEEtran}
\bibliography{IEEEabrv,reference.bib}

\end{document}